# InverSynth: Deep Estimation of Synthesizer Parameter Configurations from Audio Signals

Oren Barkan, David Tsiris, Noam Koenigstein and Ori Katz

*Abstract*— Sound synthesis is a complex field that requires domain expertise. Manual tuning of synthesizer parameters to match a specific sound can be an exhaustive task, even for experienced sound engineers. In this paper, we introduce InverSynth - an automatic method for synthesizer parameters tuning to match a given input sound. InverSynth is based on strided convolutional neural networks and is capable of inferring the synthesizer parameters configuration from the input spectrogram and even from the raw audio. The effectiveness InverSynth is demonstrated on a subtractive synthesizer with four frequency modulated oscillators, envelope generator and a gater effect. We present extensive quantitative and qualitative results that showcase the superiority InverSynth over several baselines. Furthermore, we show that the network depth is an important factor that contributes to the prediction accuracy.

*Index Terms* - deep synthesizer parameter estimation, automatic sound synthesis, inverse problems, InverSynth

## I. INTRODUCTION

THE art of sound synthesis is a challenging task traditionally reserved to sound designers and engineers [1]. Nowadays, synthesizers play a key role in electronic music production. Typically, music producers are equipped with sets of preprogrammed sound patches per synthesizer. These sets are essentially a collection of different configurations of the synthesizer parameters. Each configuration was carefully tuned by an expert to produce a different type of sound.

In the last decade, recent advances in deep learning pushed forward state of the art results in various fields such as computer vision [16], speech recognition [17], natural language understanding [18]-[23] and music information retrieval [2]-[4]. Particularly, convolutional neural networks (CNN) demonstrated extraordinary results in genre classification [5]. One of the key properties of CNNs is automatic feature learning through shared filters which enable capturing similar patterns at different locations across the signal. CNNs are a natural fit in many visual and auditory tasks and provide better generalization and overall performance [6].

In this paper, we investigate the problem of estimating the synthesizer parameter configuration that best reconstructs a source audio signal. Our model assumes that the source audio signal is generated from a similar synthesizer with hidden parameter configuration. Hence, the aim of the learning process is to reveal the hidden configuration that was used to generate the source signal. This is particularly useful in scenarios where a music artist is interested in replicating sounds that were generated by other artists / sound designers using the same synthesizer (assuming no patch is released). Indeed, one of the most frequently asked questions in music production and sound synthesis forums is: What is the recipe for replicating a sound X that was *generated* by synthesizer Y? An even more general question is: What is the recipe for replicating a sound X *using* synthesizer Y? while in the former, X is assumed to be synthesized by Y (intra-domain), in the latter, this assumption is no longer valid (cross-domain). This paper focuses on the intra-domain problem and leaves the more general cross-domain problem for future investigation. To this end, we propose the InverSynth method. InverSynth is based on strided CNNs and comes in two variants: The first variant is a spectrogram based CNN that is trained to predict the synthesizer parameter configuration from the log Short Time Fourier Transform (STFT) spectrogram [1] of the input audio signal. The second variant is capable of performing end-to-end learning directly from the raw audio to the synthesizer parameters domain, successfully. This is done by adding several convolutional layers that are designed to learn an alternative representation for the log STFT spectrogram.

The InverSynth variants are depicted in Figure 1. In Fig.1 (a), an input audio signal is transformed to a STFT spectrogram matrix, which is then fed to a CNN. The CNN analyzes the spectrogram and predicts a parameter configuration. Finally, the synthesizer is configured according to the predicted parameters values and synthesizes the output audio signal. In Fig. (b), a CNN performs end-to-end learning and predicts the parameters configuration directly from the raw audio. In addition, we compare the performance of InverSynth against two other types of fully connected (FC) neural network models: the first type is a FC network that receives a Bag of Words (BoW) representation of the spectrogram as input. The second type is a FC network that receives a set of complex hand crafted features [10] that are designed to capture spectral properties and temporal variations in the signal.

The audio signals that are used for training and testing the models are generated by synthesizer parameter configurations that are randomly sampled, i.e. for each synthesizer parameter, we define an effective range of valid values and sample the parameter value from this range, uniformly. Hence, each configuration in the dataset is obtained by a set of (parameter, sampled value) pairs.

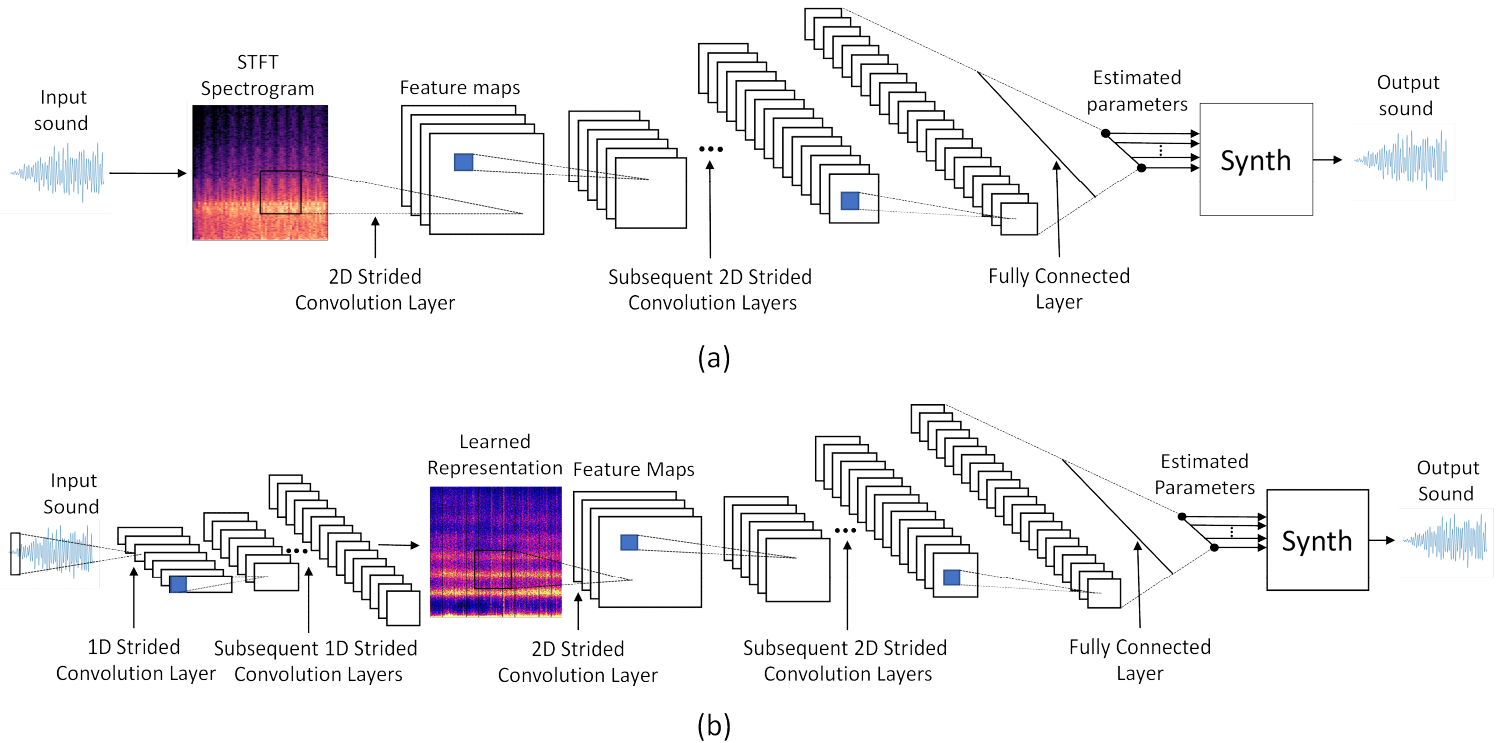

Fig. 1. The InverSynth models. (a) The STFT spectrogram of the input signal is fed into a 2D CNN that predicts the synthesizer parameter configuration. This configuration is then used to produce a sound that is similar to the input sound. (b) End-to-End learning. A CNN predicts the synthesizer parameter configuration directly from the raw audio. The first convolutional layers perform 1D convolutions that learn an alternative representation for the STFT Spectrogram. Then, a stack of 2D convolutional layers analyze the learned representation to predict the synthesizer parameter configuration.

We present a comprehensive investigation of various network architectures and demonstrate the effectiveness of InverSynth in a series of quantitative and qualitative experiments. Our findings show that the network depth is an important factor which contributes to the prediction accuracy and a spectrogram based CNN of a sufficient depth outperforms its end-to-end counterpart, while both InverSynth variants outperform FC networks and other models that utilize handcrafted features.

## II. RELATED WORK

The problem of synthesizer parameter estimation has been studied in the literature [39]. Several attempts have been made to apply traditional machine learning techniques to physical modeling of musical instruments such as bowed and plucked strings [7]–[9]. Physical modeling synthesis is mainly used for creating real-sounding instruments but is less common than subtractive synthesis with frequency modulation (FM) [1], which is one of the dominant synthesis methods in electronic music production that enables the creation of extremely diversified sounds.

Itoyama and Okuno [10] employed multiple linear regression on a set of handcrafted features for the task of learning synthesizer parameters. Our approach, differs from [10] in two main aspects: First, we solve a classification task rather than a regression task. Second, we employ deep CNN architectures with non-linear gates. CNNs have sufficient capacity to learn complex filters (weights) which enables capturing important patterns while maintaining robust generalization. This eliminates the need for handcrafted features and enables the use of STFT spectrogram or raw audio as input without further manipulations. We validate this claim, empirically, by comparing the CNN models to linear and non-linear FC models that use the hand crafted features from [10].

An additional attempt to learn synthesizer parameters through a regression was presented by M. Yee-King et al. [27,28]. They experimented with a variety of algorithms such as a hill climber, a genetic algorithm, MLP networks and RNN networks. In contrast to [27,28], we focus on network architectures that are based on strided convolutional operators. It is worth mentioning that our initial experimentations did in fact covered RNNs. However, we discovered that RNNs had major difficulty handling the lengthy audio sequences containing thousands of samples per second. Furthermore, in an effort to alleviate this difficulty, we employed convolutional layers in an attempt to compress the time axis into shorter sequences that were subsequently fed into the RNN network. Nevertheless, this approach did not yield any improvement over our final models that are presented in this paper. Finally, in contrast to both [27,28] as well as [10], we do not solve a regression task and formulate the parameter estimation task as a classification problem instead which enables better quantification and understanding of the parameter estimation accuracy.

The above papers are previously published research work dealing with the problem of synthesizer parameter estimation. Nevertheless, others have dealt with different yet somewhat related tasks. For example, Hu Yuanming, et al. [29] proposed applying a CNN with GANs using reinforcement learning to reveal the sequence of editing steps for an image, which correspond to standard retouching operations and provide some understanding of the process that it took. Sheng, Di, and György Fazekas. [30,31] employed a siamese DNN regression model to learn the characteristics of the audio dynamic range compressor (DRC). Jacovi, Alon, et-al. [32] proposed a method for end-to-end training of a base neural network that integrates calls to existing black-box functions. They showed the applications of their work in the NLP and Image domains. Yan et al. [33] proposed an automatic photo adjustment framework based on deep neural networks solving a regression problem. E-P Damskagg, et al. [34] proposed a WaveNet deep neural network that carries out a regression to virtual analog modeling and applied it to the Fender Bassman 56F-A vacuum-tube amplifier. Finally, Martinez, et al. [35,36,37] employed deep neural networks to model nonlinear audio effects. In contrast to [29]-[37], we investigate the problem of parameter estimation for sound synthesis.

III. SYNTHESIZER ARCHITECTURE AND PARAMETERS

The synthesizer architecture used in this work is implemented using JSyn [11], an open source library that provides audio synthesis API for Java. The JSyn framework was chosen for two main reasons: Most commercial synthesizers do not provide an API for generating sounds programmatically, which renders the dataset generation process impractical (in this work, we use a dataset that contains 200K instances). Second, for the sake of reproducibility, we favored an open source synthesizer over relatively expensive commercial alternatives.

Similar to most of modern synthesizers, we employ subtractive and FM synthesis [1]. The synthesizer architecture is a cascade of four components and is depicted in Fig. 2. The first component is a set of oscillators, each produces a different waveform type: sine, saw, square and triangle [1]. All oscillators are frequency modulated by a sinusoidal waveform. An oscillator function is defined as
$$y_w(f, v, A, B) = A x_w\big(2\pi f t + B\sin(2\pi v t)\big) \quad (1)$$
where $f, v, A, B$ are the carrier frequency, modulation frequency, carrier amplitude and modulation amplitude, respectively, while $w \in W$ is the waveform type with $W = \{sin, saw, tri, sqr\}$ and

$$x_{sin}(a) = \sin(a),$$
$$x_{saw}(a) = 2^{-1} - \pi^{-1}\sum_{n=1}^{\infty} n^{-1}(-1)^n \sin(na),$$
$$x_{tri}(a) = 2^{-1} - \pi^{-1}\sum_{n=1}^{\infty} n^{-2}(-1)^n \sin(na),$$
$$x_{sqr}(a) = \text{sgn}(\sin(a)).$$

Note that each oscillator $y_w$ is associated with its **own** set of $f_w, v_w, A_w, B_w$ parameters. Finally, the outputs from all oscillators are summed to $y_{osc} = \sum_{w \in W} y_w$. Therefore, the total number of parameters in the $y_{osc}$ component is 16.

The function $y_w$ from Eq. (1) is a special case of a general family of frequency modulated functions of order $n \in \mathbb{N}$:
$$y(\Psi, \Phi, \Omega) = \psi_0 x_{\phi_0}(2\pi \int_0^t \gamma_0(t_0) dt_0), \text{ with}$$
$$\gamma_0(t_0) = \omega_0 + \psi_1 x_{\phi_1}(2\pi \int_0^{t_0} \gamma_1(t_1) dt_1), \ldots, \gamma_n(t_n) = \omega_n$$
where $\Psi = (\psi_i)_{i=0}^n$, $\Phi = (\phi_i)_{i=0}^n$, $\Omega = (\omega_i)_{i=0}^n$ are the amplitudes, waveforms and frequencies series and it holds that $\forall i: \phi_i \in \Lambda$, where $\Lambda$ is a set of periodic functions symbols, i.e. the waveform symbols set. In this work, we focus on frequency modulated functions $y$ of order 1 with $\phi_0 \in W$ and $\phi_1 = sin$ as defined in Eq. (1).

The second component is the *Attack Decay Sustain Release* (ADSR) envelope generator [1] $y_{env}(x, a, d, s, r)$. This component controls the amplitude of the input $x$ at any point in the signal duration. The contour of the ADSR envelope is specified using four parameters: $a$ (*Attack*) is the time taken for initial run-up of level from zero to peak, beginning when the key is first pressed. $d$ (*Decay*) is the time taken for the subsequent run down from the attack level to the designated sustain level. $s$ (*Sustain*) is the level during the main sequence of sound's duration, until the key is released. $r$ (*Release*) is the time taken for the level to decay from the sustain level to zero after the key is released.

The third component is the filter $y_{lp}(x, f_{cut}, q)$ that consists of a low-pass filter together with a resonance [1]. Setting a cutoff frequency $f_{cut}$ ensures that all frequencies above $f_{cut}$ are cut. The resonance parameter $q$ determines a narrow band of frequencies near $f_{cut}$ that are amplified.

The last component in the chain is the gater effect that controls the rate of the amplitude. The gater is a Low Frequency Oscillator (LFO) that performs amplitude modulation to the input, according to a sine waveform with a frequency $f_{gate}$:
$$y_{gate}(x, f) = ((1 + x_{sqr}(2\pi f_{gate} t))/2) x(t).$$
Figure 2 illustrates the synthesizer function that has 23 parameters and is given by $y_{gate} * y_{lp} * y_{env} * y_{osc}$, where $*$ stands for the function composition.

IV. DATASET GENERATION

We employ the synthesizer from Section 3 to generate the dataset. Each synthesizer parameter underwent quantization to a set of 16 levels each representing a different class. Hence, we formulate the parameter estimation task as a classification problem rather than a regression - our model aims at predicting the correct class (value) for each synthesizer parameter. Through this choice, the model employs the binary cross entropy loss [6] which is easier to optimize than the L2 loss [12], especially in case of a small number of classes. Moreover, the classification formulation allows us to use several measures (Section 6) that enable better quantification and understanding of the parameter estimation accuracy.

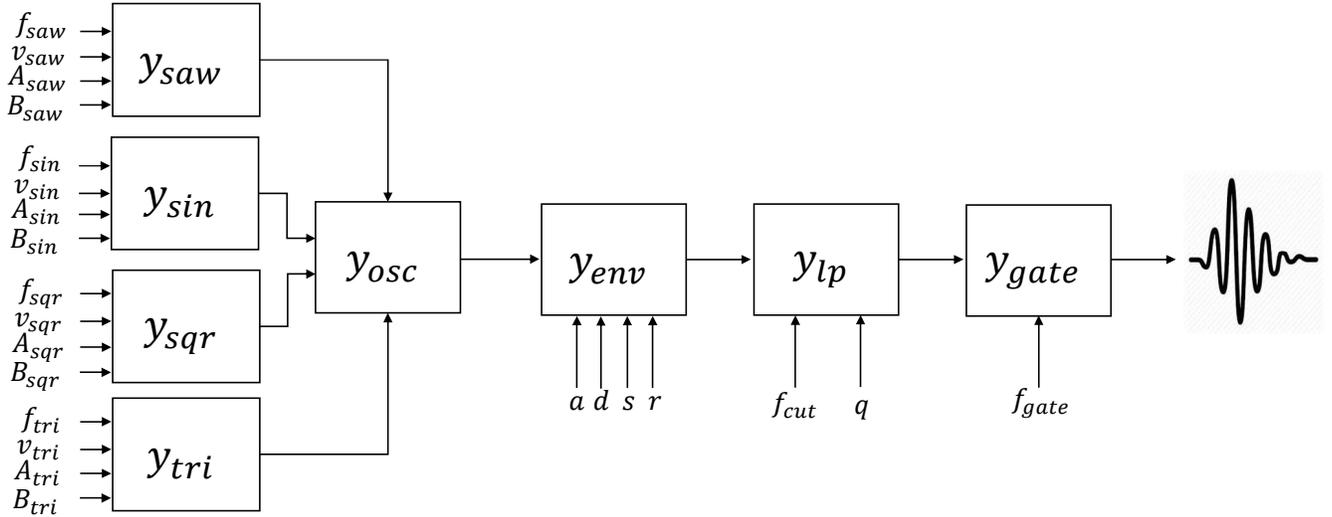

Fig. 2. The FM synthesizer used in this work. The synthesizer has four frequency modulated oscillators, ADSR envelope generator, low-pass filter and gater effect. See Section 2 for details.

The range of the carrier frequency $f$ is quantized according to $f = 2^{n/12} \times 440Hz$ with $n \in \{0..15\}$. This produces frequencies that correspond to the 16 consecutive musical notes $A_4 - C_6$. The rest of the synthesizer parameters ranges are quantized evenly to 16 classes according to the following ranges: the amplitudes and ADSR envelope parameters $A_w, a, d, s, r$ are in $[0.001, 1]$, the modulation amplitudes $B_w$ are in $[0, 1500]$ (where $B_w = 0$ means no frequency modulation), the modulation frequency, gating frequency, cutoff frequency and resonance $v_w, f_{gate}, f_{cut}, q$ are in $[1, 30], [0.5, 30], [200, 4000], [0.01, 10]$, respectively. For each parameter, the first and last classes correspond to its range limits. These ranges and sampling patterns were set to ensure a parameter quantization that is distinguishable by human hearing.

The dataset consists of $(g, h)$ pairs, where $h$ is the label vector that corresponds to a specific synthesizer parameter configuration and $g$ is the raw audio signal produced by the configuration associated with $h$. The generation process of a single pair $(g, h)$ works as follows: for each synthesizer parameter $j$, we sample a class from a uniform categorical random variable $u \in \{0, ..., 15\}$ to produce a one-hot encoding vector $h_i \in \{0,1\}^{16}$ for the sampled class. Then, we concatenate all the vectors to a single supervector $h = [h_1, ..., h_{23}] \in \{0,1\}^{368}$ consisting of one-hot encoding for each parameter in the corresponding section in $h$. Then, the synthesizer parameters are configured according to values that correspond to the sampled classes in $h$ and an output audio signal in a duration of 1 second with a sampling rate of 16384Hz $g \in [-1,1]^{16384}$ is produced.

Recall that in InverSynth, we consider two types of CNN architectures: an end-to-end (Fig. 1(b)) and a spectrogram based (Fig. 1(a)), where the latter requires further transformation over $g$. Therefore, the STFT spectrogram of $g$ is computed with half overlapping windows of size 512 to produce a matrix $S \in \mathbb{R}^{257 \times 64}$. This matrix contains 64 vectors in size of 257 that correspond to the absolute of the Fourier Transforms (FT) of 64 consecutive time windows. Note that the application of FT produces 512 complex values. However, we discard half of them since the FT of real signals is conjugate symmetric. This process is repeated 200k times to produce two datasets: $D_{E2E} = \{(g_k, h_k)\}$ and $D_{STFT} = \{(S_k, h_k)\}$ that are used for training the end-to-end and spectrogram based CNNs, respectively.

## V. INVERSYNTH AND BASELINE MODELS

We now turn to a detailed description of the participating InverSynth (Fig. 1) and other baseline models. Our goal is to learn a function from the spectrogram / raw audio domains to the synthesizer parameter domain. All models are trained using the datasets described in Section 4 to predict the correct parameter classes (values) from representations that are based either on the (log) spectrogram or raw audio. As a baseline, we further consider the replacement of the CNNs with Fully Connected (FC) neural networks. All models end with an output layer of 368 with sigmoid activations (to match $h$'s dimension). An alternative setup is to apply 23 separated softmax activations that are fed into 23 categorical cross entropy loss functions (for each synthesizer parameter in separate) and compute the final loss as the summation of the 23 loss functions. While this alternative approach seems more natural, our experiments showed that it performs on par. Therefore, we converged on using sigmoid activations with binary cross entropy loss.

In order to isolate the effect of network depth, we restrict all models to have the same number of trainable parameters, regardless of their depth. This restriction ensures that better results, when obtained using a deeper models, can be attributed to the increase in depth. We furthermore conduct an initial investigation to find a saturation point – a point in which a further increase in number of trainable parameters (model capacity) results in a marginal contribution to the model accuracy. In what follows, we describe the models used for evaluation.

Table 1. FC and HC models architectures (see Section 5.A)

| FC and FC Architectures | | | | | |
|---|---|---|---|---|---|
| FC Linear | FC1 | FC2 | FC3 | HC | HC3 |
| 1 Layer | 1 Layer | 2 Layers | 3 Layers | 1 Layer | 3 Layers |
| FC input: BoW vector in size of 1000 for FC models. HC input: Vector in size of 1000 [10] | | | | | |
| FC-869 | FC-868 | FC-603 | FC-560 | FC-869 | FC-560 |
| Drop-0.2 | Drop-0.3 | Drop-0.1 | FC-500 | Drop-0.2 | FC-500 |
| | | FC-602 | Drop-0.2 | | Drop-0.2 |
| | | Drop-0.3 | FC-400 | | FC-400 |
| | | | Drop-0.4 | | Drop-0.4 |
| FC-368(sigmoid) | | | | | |

*A. Fully Connected Neural Networks*

A FC neural network is a feed forward network in which each neuron is connected to all neurons in the previous layer [6]. This type of networks expect to have 1D input (vector). Since the training instances are the spectrogram matrices / raw audio signals, the trivial choice is to feed the FC network with the raw audio vectors / flattened spectrogram matrices. However, our initial experiments showed that both approaches produced poor results. We attribute this to the fact that both flattened spectrogram matrices and raw signals are not time invariant and of extremely high dimension (~16K).

In order to alleviate this problem, we propose to use two different types of input representation to the FC models: the first type is based on the spectrogram matrix. Specifically, we first apply PCA to the STFT frequency axis to produce a STFT-PCA matrix with a reduced frequency-PCA dimension of 64 (while retaining 97% of the variance). Then, we learn a Bag of Words (BoW) [13] representation of the STFT-PCA matrices. To this end, we Vector Quantize [13] the STFT matrices using K-means [13] with $K = 1000$ and assign each row vector in the STFT-PCA matrix to its closest centroid. This produces a count vector in size 1000, in which the $i$-th entry counts the number of vectors that were assigned to $i$-th centroid. Finally, we convert the counts to probabilities by normalizing each count vector by the sum of its entries. This results in a time invariant representation of a reduced dimensionality.

We consider four different FC model architectures that use the BoW representation as input: the first model has a single hidden layer with linear activations dubbed henceforth *FC Linear* (note that in this work, we define the number of layers in a network, as the number of hidden layers in between the input and output layers). Hence, FC Linear is equivalent to a logistic regression model with a hidden layer. The three other models are dubbed *FC1*, *FC2* and *FC3* and have 1, 2 and 3 hidden layers with ReLU [6] activations respectively. Our experimentations showed that network depths to more than three layers did not contribute to improved performance.

As an alternative input representation, we consider a set of complex hand crafted features [10] computed from the raw audio signal. These features are designed to capture spectral and temporal properties of the signal. In what follows, we briefly describe the feature extraction process. First are computed the energy levels of the signal in various bands, zero crossing rate, spectral width, spectral centroid, spectral rolloff, spectral flux, spectral peak, spectral peak and valley, spectral contrast, Mel-Frequency cepstrum Coefficients (MFCCs) and timbre of the harmonic components. These features are computed framewise and produces a 32 dimensional feature vector per frame. In order to capture temporal variation, the feature vector is further extended (by concatenation) with several types of time derivatives that produced an extended feature vector of 224 dimensions. Then, all framewise feature vectors are accumulated along the time axis via the computation of the following statistics: summation, mean, variance, skewness, minimum, maximum, median, $10^{th}$ and $90^{th}$ percentiles, where the last five are computed with their positions in time to capture temporal structure. In addition, the bottom 10 coefficients of the discrete cosine transform are computed. This procedure is repeatedly applied for 25 different segments of the signal in order to characterize the signal in different temporal regions. Finally, all segment-wise feature vectors are concatenated to produced a feature supervector in dimension of 319,200 that is reduced to dimension of 1000 by the application of PCA. The reader is referred to Section 2 in [10] for a detailed description of the exact feature extraction process.

We consider two different FC model architectures that use the input representation from [10]. The first is a linear FC model that is equivalent to the one from [10], but uses a classification output layer instead of regression (In our initial experiments, we tested regression models and observed they perform worse than classification models). We dub this model 'HC' (hand crafted). In addition, we investigated whether a nonlinear FC model can benefit from using the features from [10]. We found that a FC network with 3 ReLU activated layers and a dropout in between achieves the largest improvement over HC [10], while further increase in depth or number of trainable parameters results in overfitting with worse values of validation loss. We dub this model 'HC3'. Note that HC3 is an improvement we suggest, in order to showcase the potential gain that can be achieved by using the features from [10] with deep neural networks.

To minimize the risk of overfitting, all models in this section employ a dropout [6] after each hidden layer. The exact architectures, for each FC model is detailed in Table 1, where FC-k stands for a FC layer with output size of k and Drop-p stands for a dropout that is applied to the input of the next layer with a probability *p*. All FC models have ~1.2M trainable parameters.

Table 2. InverSynth models architectures (See Section 5.B for details)

| InverSynth Architectures | | | | | | | |
|---|---|---|---|---|---|---|---|
| Conv 1 | Conv2 | Conv3 | Conv4 | Conv5 | Conv6 | Conv6XL | ConvE2E |
| 2 Layers | 3 Layers | 4 Layers | 5 Layers | 6 Layers | 7 Layers | 7 Layers | 11 Layers |
| Input (64 X 257 STFT Spectrogram) | | | | | | | Input (16384 raw audio) |
| C(38,13,26,13,26) | C(35,6,7,5,6) | C(32,4,5,3,4) | C(32,3,4,2,3) | C(32,3,3,2,2) | C(32,3,3,2,2) | C(64,3,3,2,2) | C(96,1,64,1,4) |
| | C(87,6,9,5,8) | C(98,4,6,3,5) | C(65,3,4,2,3) | C(98,3,3,2,2) | C(71,3,3,2,2) | C(128,3,3,2,2) | C(96,1,32,1,4) |
| | | C(128,4,6,3,5) | C(105,3,4,2,3) | C(128,3,4,2,3) | C(128,3,4,2,3) | C(128,3,4,2,3) | C(128,1,16,1,4) |
| | | | C(128,4,5,3,4) | C(128,3,5,2,4) | C(128,3,3,2,2) | C(128,3,3,2,2) | C(257,1,8,1,4) |
| | | | | C(128,3,3,2,2) | C(128,3,3,2,2) | C(256,3,3,2,2) | C(32,3,3,2,2) |
| | | | | | C(128,3,3,1,2) | C(256,3,3,1,2) | C(71,3,3,2,2) |
| | | | | | | | C(128,3,4,2,3) |
| | | | | | | | C(128,3,3,2,2) |
| | | | | | | | C(128,3,3,2,2) |
| | | | | | | | C(128,3,3,1,2) |
| FC-512 | | | | | | | |
| FC-368(sigmoid) | | | | | | | |

*B. The InverSynth Models*

A CNN [6] is a special type of neural network that employs weight sharing. This property enables the reuse of the same set of weights in all positions in the input. The CNN weights act as local filters that are being convolved with the input. This results in a network with a fewer neuron connections than in FC networks (in which every neuron is connected to all neurons in the previous layer). In addition, Max-Pooling (MP) [6] is often applied to downsample the input in between layers.

InverSynth uses 2D CNNs as these were found to improve on the state of the art in music and audio related tasks [5]. Different from [5], we do not perform any type of pooling operations. Instead, InverSynth uses strided convolutional layers [6], as we found this approach to significantly outperform the traditional setup of ordinary convolutions layers with pooling in between.

We consider to variants of InverSynth: the first variant is a spectrogram based CNN that receives the (log) spectrogram matrix as input. This network learns filters that analyze the input in both frequency and time axes, simultaneously and is illustrated in Fig. 1(a). We investigate seven different spectrogram based CNNs. The first six InverSynth models share the same number of 1.2M trainable parameters, but vary by network depth. These models are dubbed Conv1,…,Conv6 and have 1,…,6 2D strided convolutional layers, respectively. The seventh CNN is dubbed Conv6XL and has 6 strided convolutional layers, but 2.3M trainable parameters. The reason we further include Conv6XL in the evaluation is to check whether increasing the model capacity, in terms of number of trainable parameters, further contributes to the prediction accuracy. Finally, it is worth noting that in our initial experimentation, the use of more than 6 convolutional layers did not materialize to any further improvements.

The second variant of InverSynth is an end-to-end CNN that receives the raw audio signal as input. This model is dubbed ConvE2E and further aims at learning a set of filters that produce a transformation on the raw audio, which is an *alternative* to the log STFT spectrogram. Hence, the first four layers in ConvE2E are dedicated to transform the 16K dimensional input signal to a matrix in the exact same size of the STFT matrix (64x257). These four layers are 1D strided convolutional layers that operates on the time axis only. This is followed by additional six 2D strided convolutional layers that are identical to those of Conv6 model. Due to the extra four layers, ConvE2E has 1.9M trainable parameters. The ConvE2E architecture is illustrated in Fig. 1(b). Finally, all CNN models have an additional FC hidden layer in between the last convolutional layer and the output layer.

The exact InverSynth models architectures are detailed in Table 2, where C(F,K1,K2,S1,S2) stands for a ReLU activated 2D strided convolutional layer with F filters in size of (K1,K2) and strides (S1,S2). In the case of ConvE2E, the first four layers degenerates to 1D strided convolutions by setting both K1 and S1 to 1. Finally, it is worth noting that no dropout is applied in the CNNs, since we found that CNNs less tend to overfitting.

VI. EXPERIMENTAL SETUP AND RESULTS

In what follows, we describe the experimental setup, evaluation measures and present quantitative and qualitative results. The reported results are obtained using 10 fold cross validation on the datasets described in Section 4.

The evaluated models are specified in Tables 1 and 2. All models are optimized w.r.t. the binary cross entropy loss [6] using the Adam [14] optimizer with a minibatch size of 16 for 100 epochs. The best weights for each model were set by employing an early stopping procedure. We observed that the early stopping procedure stopped the training before 100 epochs for all models and we used the best model for each model type. Nevertheless, we continued the training in order to produce a coherent plot in which the progression of both training and validation losses for all models appears over the entire X-axis.

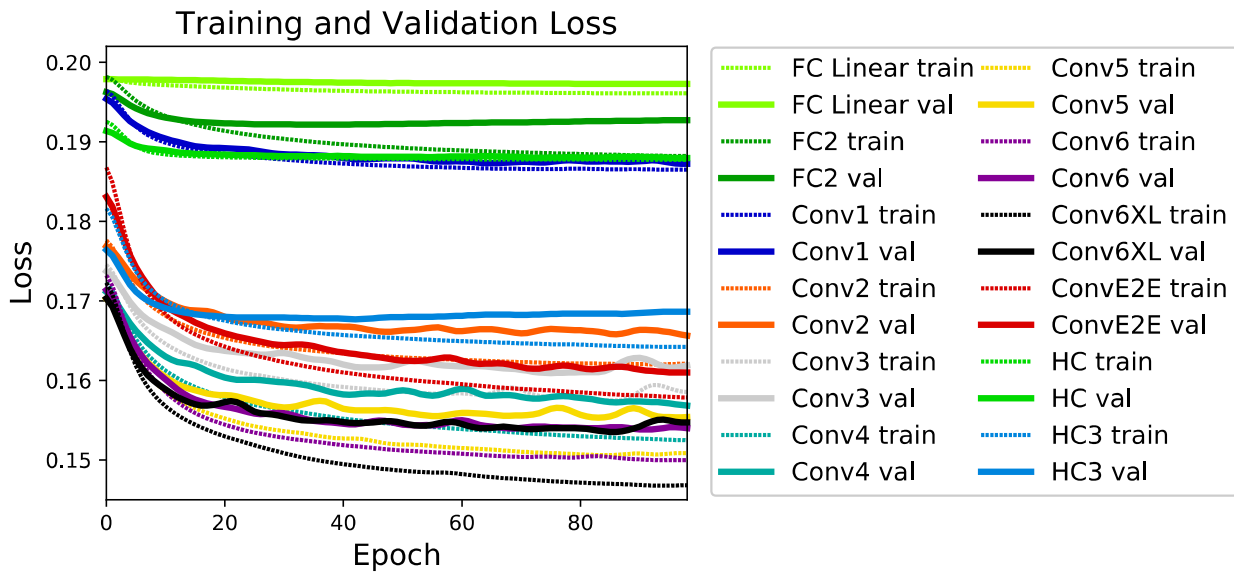

Fig. 3. Training and validation loss graphs for all models.

*A. Quantitative Results*

The training and validation loss per model are displayed in Figure 3. As can be seen in Fig. 3, for all models, both training and validation losses saturate on epoch 100. Specifically, HC3 seems to suffer from a very mild overfitting, but the early stopping procedure ensures the best HC3 model is used (which is the model from epoch 24). We can see that most of the InverSynth models obtain significantly lower validation loss values than their FC and HC counterparts. Specifically, Conv6XL and Conv6 exhibit best results which is in par with each other. This finding indicates that increasing the number of trainable parameters from 1.2M to 2.3M has a negligible effect for CNNs of a sufficient depth. The most significant gain (lower loss) is obtained when moving from Conv1 to Conv2 and then from Conv2 to Conv3. We attribute this to the fact that Conv1 and Conv2 use large stride values which result in an over-aggressive subsampling. The ConvE2E model underperforms Conv4, Conv5, Conv6 and Conv6XL, but is on par with Conv3 and outperforms all FC and HC models. Hence, we conclude that the STFT spectrogram contains further crucial information that ConvE2E fails to extract from the raw audio. Nevertheless, ConvE2E manages to learn filters that are better than the complex handcrafted features from [10], even when combined with a deep FC network (HC3).

When we turn our attention to the FC model we notice that FC2 outperforms all other FC models, yet it still very much behind all the CNNs models. FC Linear exhibits worst results among all models. Note that we omit the loss graphs of FC1 and FC3 for better clarity as these almost completely overlap with FC2. Examining the performance of the HC models, we observe the following trends: HC which is a linear model outperform all non-linear FC models and is on par with Conv1. This can be explained by the fact that HC uses hand crafted features that are far more complex than the STFT spectrogram and its BoW representation. HC3 significantly outperform HC and Conv1. However, HC3 is still outperformed by most of the Conv models including ConvE2E.

Figure 3 provide further indication that network depth plays an important factor: the deeper the network the better it manages to learn (recall that all models have the same number of trainable parameters except for Conv6XL and ConvE2E). Specifically, we observe that a large reduction in loss is obtained by moving from HC to HC3, despite the fact that both models have the same number of trainable parameters. This is another evidence for the importance of network depth which results in additional nonlinear transformations. However, the contribution from depth becomes marginal, when using more than 5, 2 and 3 layers for InverSynth, FC and HC models, respectively.

In terms of generalization, FC1, FC2, FC3 and HC3 appear to start overfitting after ~20 epochs. We tried to alleviate this by increasing the dropout values and applying L2 regularization, but this attempt failed to yield better loss values. In the case of the InverSynth models, no overfitting is observed, despite the fact that no regularization is applied. We believe that this is due to the nature of CNNs that enable weight sharing, as well as the fact that the input signals originate in the stationarity periodic functions family (Eq. (1)) (excluding the ADSR envelope effect).

Although the models are optimized w.r.t. the binary cross entropy loss, our goal is to measure the reconstruction quality in terms of human hearing. As we are not aware of any quantitative measure that correlates well with human hearing, we propose several additional measures to evaluate the models performance in multiple resolutions: accuracy per parameter and group of parameters, reconstruction quality in both frequency and time-frequency domains.

**Table 3.** MPR values for each combination of model and parameter (higher is better). The last row depicts the mean MPR across all models. The best MPR values for each parameter is highlighted in each row. The Conv6XL model "wins" on most parameters and yields results that are very close to the winning model on most of the other parameters.

| Parameter | FC Linear | FC1 | FC2 | FC3 | Conv1 | Conv2 | Conv3 | Conv4 | Conv5 | Conv6 | Conv6XL | ConvE2E | HC | HC3 |
|---|---|---|---|---|---|---|---|---|---|---|---|---|---|---|
| $a$ | 59.34 | 61.46 | 61.59 | 61.23 | 69.15 | 69.28 | 69.67 | 69.91 | 70.66 | 71.28 | 70.85 | **84.38** | 75.74 | 83.59 |
| $d$ | 53.23 | 52.99 | 53.04 | 53.2 | 53.3 | 53.7 | 53.69 | **54.05** | 53.64 | 53.53 | 53.53 | 53.98 | 53.87 | 53.12 |
| $s$ | 53.09 | 53.48 | 53.2 | 53.45 | 52.76 | 53.69 | 53.38 | 53.43 | 53.55 | 53.32 | 53.55 | **54.01** | 53.33 | 53.09 |
| $r$ | 59.49 | 61.76 | 62.25 | 61.99 | 81.32 | 83.79 | 85.88 | 88.04 | 88.43 | **90.47** | 90.23 | 88.52 | 76.4 | 88.46 |
| $f_{gate}$ | 54.49 | 55.48 | 55.89 | 55.49 | 81.03 | 96.02 | 96.6 | 97.66 | 97.85 | 97.54 | **97.88** | 92.81 | 91.99 | 94.02 |
| $f_{cut}$ | 91.27 | 92.83 | 93.05 | 92.9 | 98.64 | 99.07 | 99.48 | 99.59 | 99.64 | **99.71** | 99.68 | 98.5 | 96.76 | 99.27 |
| $q$ | 79.84 | 82.22 | 82.59 | 82.38 | 90.46 | 92.55 | 94.39 | 95.34 | 96.1 | 96.46 | **96.62** | 93.3 | 85.25 | 94.55 |
| $A_{saw}$ | 65.41 | 66.99 | 67.96 | 68.09 | 68.05 | 72.05 | 74.02 | 74.95 | 76.36 | 76.5 | **77.12** | 69.54 | 66.54 | 75.66 |
| $B_{saw}$ | 82.58 | 83.25 | 82.8 | 82.46 | 81.03 | 83.75 | 84.04 | 85.01 | 85.4 | 85.7 | **85.73** | 79.76 | 81.97 | 81.85 |
| $v_{saw}$ | 55.31 | 56.04 | 56.51 | 56.28 | 59.62 | 66.56 | 67.12 | **67.98** | 67.33 | 67.18 | 66.79 | 65.58 | 64.46 | 63.75 |
| $f_{saw}$ | 72.54 | 74.02 | 73.36 | 72.71 | 70.31 | 77.74 | 78.27 | 79.93 | 79.65 | 79.86 | **80.16** | 73.85 | 73.72 | 74.93 |
| $A_{sin}$ | 54.56 | 55.52 | 56.05 | 55.91 | 59.73 | 63.33 | 65.3 | 65.84 | 66.45 | 66.8 | **67.28** | 64.7 | 60.16 | 65.79 |
| $B_{sin}$ | 80.43 | 81.03 | 81.33 | 80.74 | 82.06 | 84.3 | 84.43 | 85.67 | 85.95 | 86.42 | **86.5** | 81.49 | 83.72 | 83.89 |
| $v_{sin}$ | 54.24 | 54.87 | 54.97 | 54.48 | 60.91 | 67.88 | 68.18 | **68.73** | 68.53 | 68.38 | 68.02 | 67.45 | 65.4 | 64.4 |
| $f_{sin}$ | 60.19 | 62.09 | 62.12 | 61.15 | 64.73 | 79.08 | 79.5 | 80 | 80.62 | **80.55** | 80.41 | 75.33 | 73.26 | 75.23 |
| $A_{sqr}$ | 65.17 | 67.22 | 68.13 | 68.07 | 65.51 | 71.51 | 73.04 | 74.46 | 75.31 | 75.65 | **76.46** | 72.08 | 65.88 | 75.69 |
| $B_{sqr}$ | 85.98 | 87.61 | 87.6 | 87.34 | 85.65 | 88.54 | 89.01 | 89.99 | 90.51 | 90.77 | **90.85** | 83.89 | 86.69 | 87.98 |
| $v_{sqr}$ | 57.42 | 59.31 | 59.8 | 59.76 | 63.89 | 69.47 | 69.94 | **70.25** | 69.99 | 69.68 | 69.55 | 68.91 | 68.39 | 67.82 |
| $f_{sqr}$ | 76.29 | 78.76 | 78.44 | 77.79 | 75.02 | 83.2 | 83.88 | 84.8 | 85.39 | 85.79 | **86.27** | 78.38 | 78.46 | 81.62 |
| $A_{tri}$ | 54.15 | 54.93 | 55.39 | 55.38 | 58.53 | 61.78 | 63.1 | 63.99 | 64.76 | 64.7 | **65.41** | 61.73 | 57.55 | 61.61 |
| $B_{tri}$ | 79.8 | 80.41 | 80.48 | 79.97 | 81.15 | 82.43 | 82.53 | 83.76 | 84.27 | **84.45** | 84.27 | 80.41 | 81.99 | 82.15 |
| $v_{tri}$ | 54.28 | 54.49 | 54.48 | 54.3 | 59.43 | 66.55 | 66.94 | **67.49** | 67.25 | 66.91 | 66.61 | 66.8 | 63.49 | 63.07 |
| $f_{tri}$ | 59.53 | 60.96 | 61.12 | 59.52 | 63.61 | 76.66 | 77.3 | 78.13 | **78.48** | 78.03 | 77.9 | 73.62 | 69.94 | 72.71 |
| **Mean MPR** | 65.59 | 66.85 | 67.05 | 66.72 | 70.69 | 75.78 | 76.51 | 77.35 | 77.66 | 77.81 | **77.9** | 76.17 | 72.8 | 75.8 |

*1) Mean Percentile Rank (MPR) based evaluation*

The first evaluation measure is the Mean Percentile Rank (MPR) which is computed per synthesizer parameter. Formally, we denote by $r_i$ the ranked position of the correct class, when measured against the other classes based on prediction scores output by the model. In our case, we have 16 classes. Hence, the *MPR* measure is computed according to $MPR = 100 \times (1 - \frac{1}{|\mathcal{T}|} \sum_{i \in \mathcal{T}} \frac{r_i}{15})$ where $\mathcal{T}$ is the number of test instances. Note that $0 \leq MPR \leq 100$, where $MPR = 100$ is the optimal value and $MPR = 50$ can be achieved by random predictions.

Table 3 presents the obtained MPR values for each combination of model and synthesizer parameter. The trends from Figure 3 seem to concur with those of Table 3: Generally, the InverSynth models exhibit better MPR values than FC models. An exception is the MPR values obtained for the decay and sustain parameters $d, s$. This can be explained by the fact that these parameters have a negligible influence on the signal amplitude comparing to the attack and release. Moreover, the ADSR envelope determines the change in amplitude over time and has a small effect over the sound timbre. Therefore, we find that bad MPR values for $d, s$ do not mean bad signal reconstruction in terms of human hearing.

An interesting observation is that ConvE2E, HC and HC3 models produce significantly better MPR scores for the attack parameter $a$. This implies that ConvE2E dedicates filters for capturing volume transients that are also captured by the temporal feature extraction scheme of [10]. In addition, we observe that for the modulation frequency $v_w$, the best MPR values are obtained by the Conv4 across all waveforms $W$, but by a small margin over Conv5 / Conv6. The last row in Table 3 contains the mean MPR values per model across all synthesizer parameters. The Conv6XL model exhibits the best mean value followed by the Conv6 model. ConvE2E is on par with Conv3 and slightly better than HC3. Finally, HC models significantly outperform all FC models.

*2) Top-k mean accuracy based evaluation*

The second measure is the top-k mean accuracy. For a given test example, the top-k accuracy function outputs 1 if the correct class rank is among the top k predicted classes by the model and 0 otherwise. The top-k mean accuracy is obtained by computing the top-k accuracy for each test example and then taking the mean across all examples. In the same manner as done in the MPR analysis, we compute the top-k mean accuracy per synthesizer parameter for $k = 1, ..., 5$.

In order to inspect the models performance w.r.t. each functional in the synthesizer, we group the synthesizer parameters according to the following functionality groups: *Filter* - contains the parameter $f_{cut}, q$ . *Notes*, *Amplitude LFO*, *Frequency LFO* and *Oscillators Amplitude* - contain the parameters $f_w, B_w, v_w, A_w$ for all $w$, respectively. The last two groups are the *Amplitude ADSR* group that contains the parameters $a, d, s, r$ and the *All* group that contains all the synthesizer parameters. For each combination of model and parameter group we compute the mean of the top-k mean accuracy across all parameters in that group and for all k values.

Figure 4 break down the top-k mean accuracy as a function of k for each combination of parameter group and model. The results in Fig. 4 correlate well with those of Tab. 3. The best accuracy values are obtained for the *Filter* and *Amplitude LFO* groups, where the margin between InverSynth and FC models is most significant. Furthermore, in all graphs the InverSynth models significantly outperform the FC models. Additionally, we observe that for the *Amplitude ADSR* parameter group, ConvE2E produces the best accuracy graph followed by the *HC3* group. This result is attributed to the ability of these models to predict the attack parameter significantly better than the other models. Finally, an interesting pattern is observed in the *Notes* graph: while ConvE2E significantly underperforms Conv4 – Conv6XL for k = 1, it becomes the champion for k > 1. This means that ConvE2E manages to infer a better ranking of musical notes, when considering the predictions beyond the one at the top.

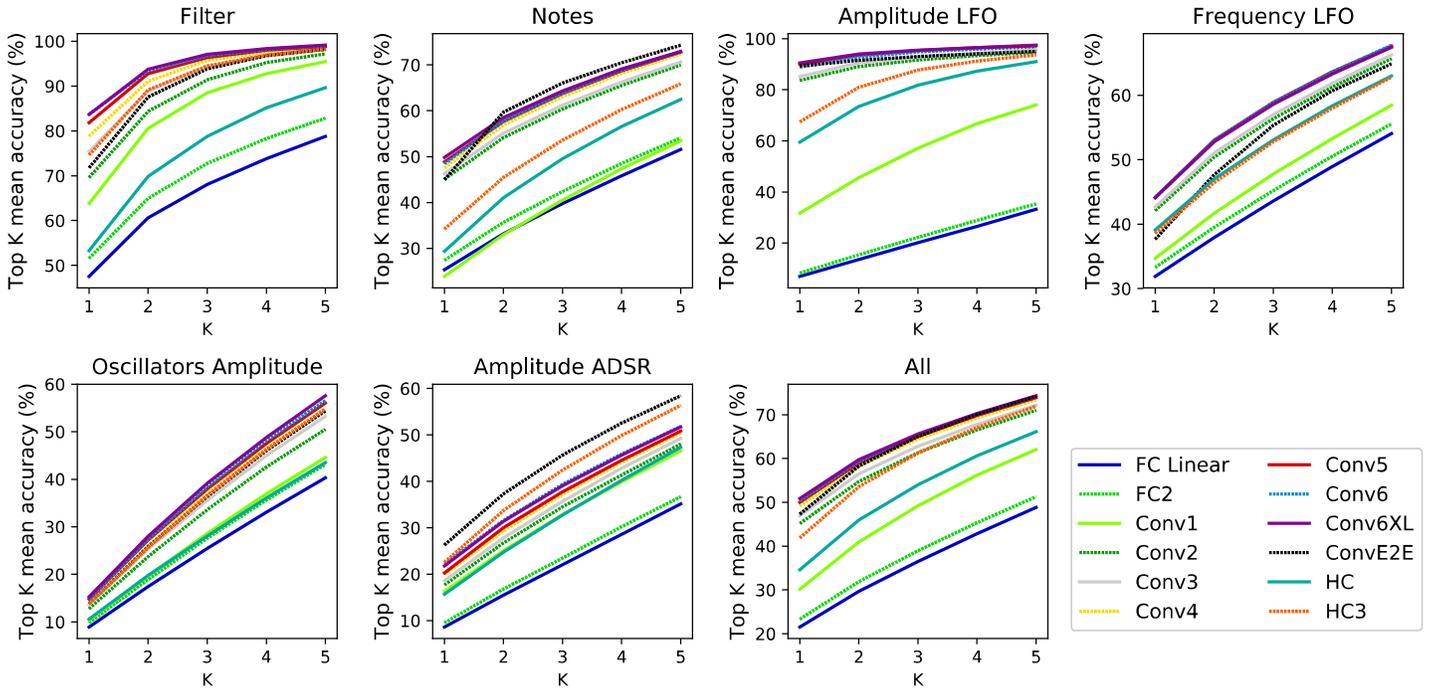

Fig. 4. Top-k mean accuracy graphs for various combinations of models and parameter groups. See Section 6.A.2 for details.

*3) Mean Absolute Error based evaluation*

The third evaluation measure is the Mean Absolute Error (MAE). The motivation for using MAE is due to the ordinal structure that exist in the 16 classes of every synthesizer parameter. For example, assume the correct class is class 8, then predicting class 9 or 7 is much better than predicting class 1 or 16. In other words, we wish to investigate the distance between the true and predicted classes. The MAE measure reveals a deeper analysis with respect to this distance.

Figure 5 presents the MAE (lower is better) scores for each synthesizer parameter. We focus on five models that best perform among the different family of models (Conv6XL, Conv6, Conv5, ConvE2E, and the HC3 model which is the leading FC model). We can see that for the attack parameter

ConvE2E and HC3 produce the best results. This correlates well with Tab. 3. In addition, ConvE2E outperforms the other models on the decay and release parameters. For the rest of parameters, Conv6XL and Conv6 are the champions.

Figure 6 presents the MAE analysis for the same parameter groups as in Fig. 4. Noticeably, the trends from Fig. 4 repeat in Fig. 6: The best performing models are Conv6XL and Conv6. The best scores are obtained for the *Filter* group followed by the *Amplitude LFO* group. The *ADSR Amplitude* group exhibit the worst performance, but within the group, the ConvE2E model is a clear winner. We conclude that despite the advantage of spectrograms-based CNN in most synthesizer parameter estimation, it still lags behind ConvE2E on the amplitude envelope estimations. We conjecture this is due to the fact that ConvE2E learns to estimate the ADSR parameters from the raw audio in an end-to-end fashion. Lastly, we observe that both InverSynth variants outperform the HC3 model, which employs a FC network to analyze a complex set of handcrafted features.

*4) Spectral reconstruction quality evaluation*

In Sections 5.A.1 and 5.A.2, we evaluated the parameter estimation accuracy obtained by the models. In this section, we investigate how well the pipeline as a whole reconstructs the spectral properties of the input signal using the different models. We focus the subset of best performing models from each model type (Linear FC, non-linear FCs, HCs and InverSynth models).

The first two evaluations are performed in the STFT domain. For each test set signal $x_i$, we compute its (log) STFT spectrogram $S_i$. Then, $S_i$ is fed to the model to produce the synthesizer parameter configuration, which is then used to

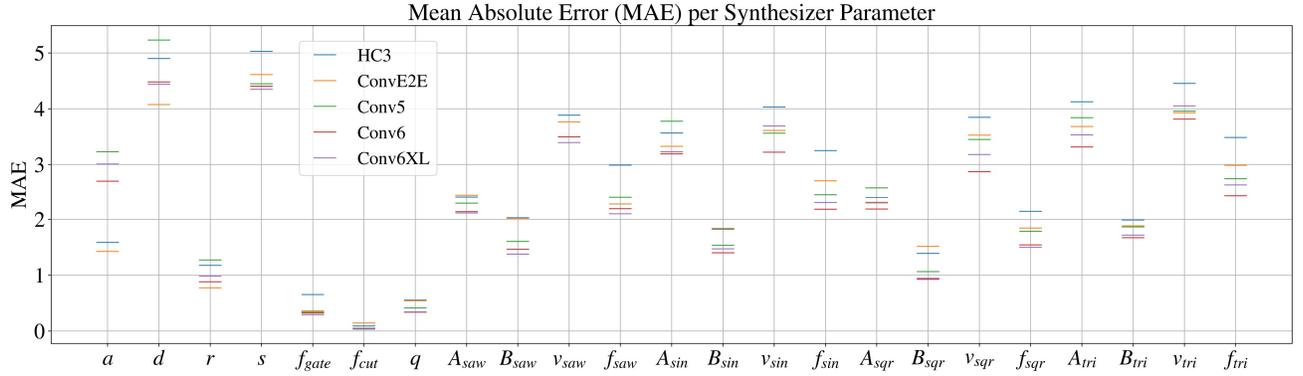

Fig. 5. Mean Average Error (MAE) for various combinations of parameters and models. See Section 6.A.3 for details.

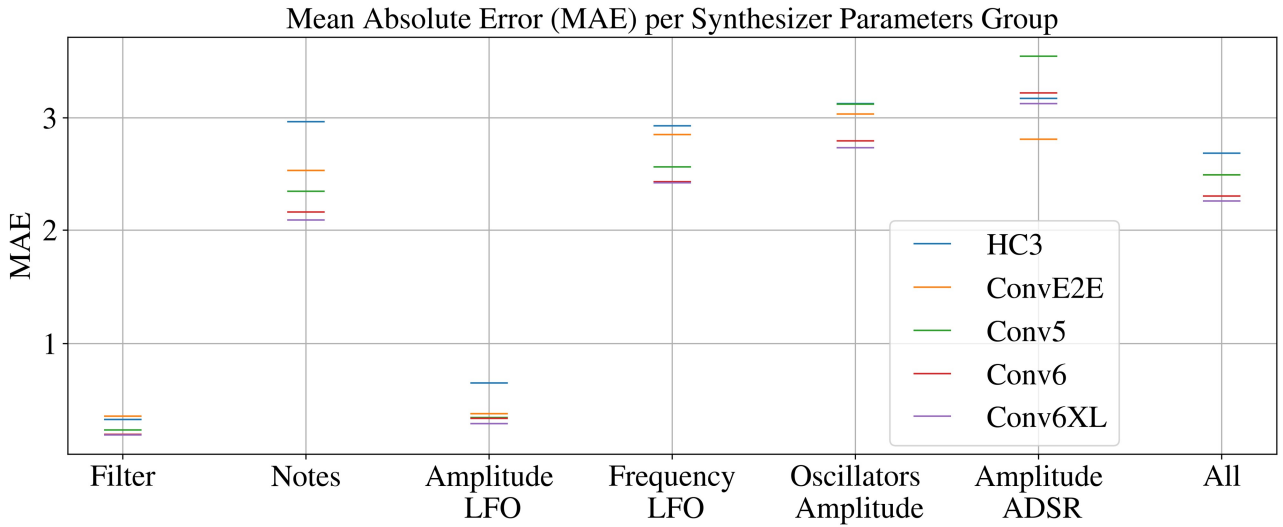

Fig. 6. Mean Average Error (MAE) for various combinations of parameters group and models. See Section 6.A.3 for details.

Table 4. Reconstruction quality for the evaluated models in time and frequency domains. $\rho$ values are x100. See Section 6.A.3 for details.

| Measure/Model | Conv6XL | Conv6 | Conv5 | Conv4 | ConvE2E | HC3 | HC [10] | FC2 | FC Linear |
|---|---|---|---|---|---|---|---|---|---|
| $\rho$ mean (STFT) | **92.04** | 91.38 | 91.09 | 90.97 | 88.33 | 88.04 | 84.62 | 77.22 | 75.24 |
| $\rho$ median (STFT) | **95.61** | 95.39 | 95.1 | 94.99 | 93.41 | 92.59 | 88.53 | 82.32 | 80.83 |
| $\rho$ mean (FT) | **76.13** | 74.28 | 73.46 | 73.29 | 71.36 | 64.39 | 59.68 | 58.18 | 54.74 |
| $\rho$ median (FT) | **90.22** | 88 | 87.4 | 86.29 | 80.61 | 68.07 | 59.73 | 57.19 | 51.15 |
| $F_\Delta$ mean (STFT) | **708.94** | 740.32 | 757.69 | 765.55 | 904.12 | 918.78 | 1117.53 | 1336.7 | 1427.03 |
| $F_\Delta$ median (STFT) | **679.36** | 704.99 | 726.87 | 743.97 | 884.96 | 916.52 | 1084.59 | 1267.87 | 1328.79 |

synthesize an output signal $x_o$ using the synthesizer. Finally, we compute the spectrogram $S_o$ of $x_o$. Note that in the case of ConvE2E and HC models $x_i$ is fed as input (HC models further require the preprocessing of [10] on $x_i$)

We propose two measures to quantify the reconstruction quality in the log spectrogram domain. The first one is the Frobenius norm of the difference between $S_i$ and $S_o$ (log-spectral distance). To this end, we denote $F_\Delta = \sqrt{Tr(UU^T)}$ with $U = S_i - S_o$. Lower $F_\Delta$ values indicate better reconstruction quality.

The second measure is the Pearson Correlation Coefficient (PCC) $\rho(x,y) = \frac{cov(x,y)}{\sigma_x \sigma_y}$. Since $\rho$ is defined over vectors rather than matrices, we flatten $S_i$ and $S_o$ to vectors by performing rows concatenation and then compute their PCC value.

The third evaluation measure is designed to quantify the reconstruction quality in the Fourier domain. To this end, we compute the PCC over the absolute of the Fourier Transforms (FT) of $x_i$ and $x_o$. Different from the STFT, which provides the frequency information per short time frames, the FT provides a global representation that aggregates the frequency information from the entire signal.

Table 4 presents the mean and median values of $\rho$ (x100) $F_\Delta$

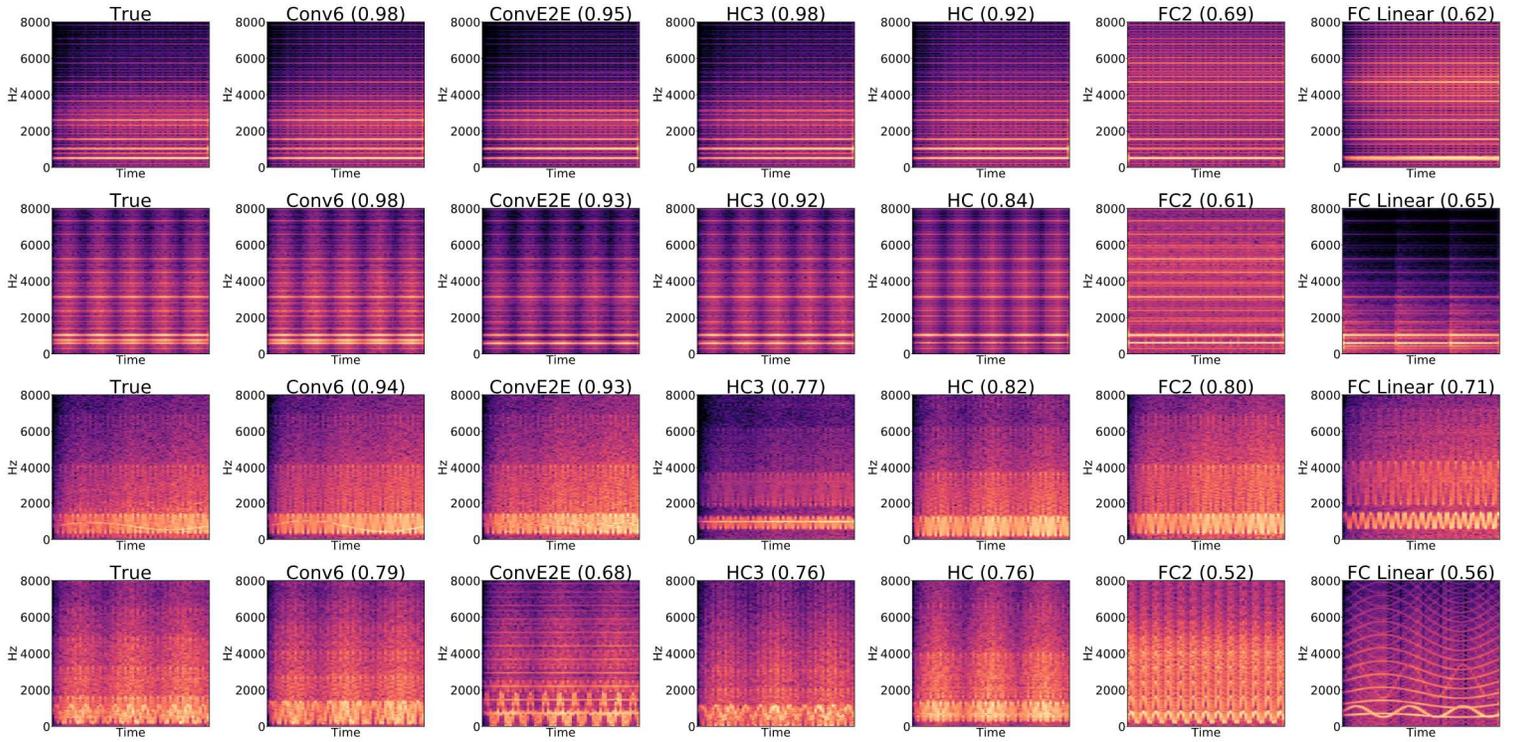

Fig. 7. Four spectrogram samples and their reconstructions. The left most spectrogram in each row is the original signal followed by reconstructed signals using different models. The PCC value in the STFT domain is depicted next to each model.

that were computed over the test set for the STFT and FT domains, accordingly. We can see that the same trend exists in all measures – spectrogram based CNNs perform the best by the network's depth. ConvE2E outperforms the HC3 and HC models (their performance is also ordered by depth). FC models perform the worst. Hence, we conclude that network depth is a key contributor to the reconstruction quality. We further observe that PCC values obtained in the FT domain are significantly worse than PCC values obtained in STFT domain. This might be explained by the fact that the models are provided with the STFT spectrogram in which the phase information is lost. Hence, we believe that by including the phase information to the models, an improvement in FT domain PCC values can be expected.

### B. Qualitative Results

In this section, we present several examples that demonstrate the effectiveness of InverSynth. All examples were chosen randomly from the test set. The audio files of these examples are available online [15].

Figure 7 presents four examples. Each row corresponds to a different example and contains (left to right): the true spectrogram of the signal and the spectrograms obtained for the reconstructed signals using Conv6, ConvE2E, HC3, HC, FC2 and FC Linear models. In addition, for each reconstruction, we report the PCC value in the STFT domain. In what follows, we discuss each example (top to bottom).

The first example is of a signal with no frequency modulation and a mild amount of gating effect. We can see that the InverSynth and HC models produce parameter configurations that result in high quality reconstruction (both visually and in terms of PCC values). This is in contrast to the FC models that produce poor results: both reconstructions contain emphasized frequency bands that do not exist in the original signal. Moreover, the FC Linear model misses a dominating low frequency band that exists in the original signal.

The second example is of a signal with no frequency modulation, but with a decent amount of gating effect. In this example, Conv6 produces a nearly perfect reconstruction, while ConvE2E misses some of the low frequency bands and exhibits an attenuated reconstruction. HC models seem to miss the same low frequency bands but exhibit less attenuation in high frequency bands. The FC2 model completely fails to recognize the gating effect and produces a reconstruction with emphasized frequency bands that do not exist in the original signal. The FC Linear model produces a wrong gating pattern and a severe attenuation of high frequency bands w.r.t the original signal.

The third example is of a frequency modulated signal with no gating. We can see that both Conv6 and ConvE2E produce decent reconstructions, while ConvE2E misses some of the FM pattern. HC3 produces a noticeably worse reconstruction than HC and FC2, while FC Linear exhibits the worst reconstruction values.

The last example is of a frequency modulated signal with a mild gating effect. We can see that Conv6 obtains a reconstruction that is the closest to the original spectrogram (visually), followed by the HCs models, while the reconstructions given by ConvE2E and FC models are poor.

This limited set of sample exemplifies earlier results that indicate that both InverSynth variants outperform other types of models.

**Table 5.** Mean Opinion Scores (MOS) results (Section 6.C).

| Conv6XL | Conv6 | ConvE2E | HC3 | FC2 |
|---|---|---|---|---|
| 4.72 | 4.81 | 4.36 | 3.18 | 1.9 |

*C. Human evaluation*

In this section, we present a subjective evaluation that is based on human judgement. The main goal of this research is to produce a synthesizer parameter configuration that best reconstruct the original source sound. Hence, the ultimate measure for the reconstruction quality is the human perception. We therefore performed a listening test and report Mean Opinion Score (MOS) [38]. In this test, a user is required to rate the reconstructed sound w.r.t. the original sound in a 1-5 discrete score scale. The listening test is based on 9 users that rate reconstructed signals for 10 different samples produced by the FC2, HC3, ConvE2E, Conv6 and Conv6XL models. The MOS is computed by averaging the user ratings for each combination of model and a sound sample.

Table 5 presents the MOS results. The clear winners are the InverSynth models, Conv6XL and Conv6 that perform on par followed by the InverSynth ConvE2E model. HC3 perform significantly worse than the InverSynth models and FC2 performs the worst. These results support the empirical findings from Sections 6.A and 6.B and indicate a correlation between both parameter estimation, spectral reconstruction measures and actual human perception.

## VII. CONCLUSION

This paper proposes InverSynth - a method for predicting synthesizer parameter values that best reconstruct a given audio signal. Synthesizer parameter values are quantified into discrete values and the task is formulated as classification problem (rather than regression). The predicted parameters lead to a reconstructed audio signal that can be compared to the source signal.

Different variant of InverSynth models are investigated and evaluated against other baseline methods, quantitatively and qualitatively. These experimentations show that a spectrogram based InverSynth models and an end-to-end InverSynth model (that analyzes the raw audio signal) outperform FC and HC models (with HC models outperforms FC models). Furthermore, network depth is investigated and found to be an important factor that contributes to the prediction accuracy.

In the future, we plan to extend the evaluations to the cross-domain scenario and investigate very deep models [24, 25]. We further plan to integrate a proxy to the synthesizer function into the backpropagation process in order to train models to reconstruct the (log) STFT directly. In addition, we plan to evaluate the same methodology on a synthesizer function that incorporates higher order of modulations with different types of modulators.

REFERENCES

[1] M. Russ, Sound Synthesis and Sampling. 2009.

[2] E. J. Humphrey, J. P. Bello, and Y. LeCun, "Moving Beyond Feature Design: Deep Architectures and Automatic Feature Learning in Music Informatics," International Society for Music Information Retrieval Conference (ISMIR), no. Ismir, pp. 403–408, 2012.

[3] E. J. Humphrey, J. P. Bello, and Y. Lecun, "Feature learning and deep architectures: New directions for music informatics," Journal of Intelligent Information Systems, vol. 41, no. 3, pp. 461–481, 2013.

[4] S. Dieleman and B. Schrauwen, "End-to-end learning for music audio," in ICASSP, IEEE International Conference on Acoustics, Speech and Signal Processing - Proceedings, 2014, pp. 6964–6968.

[5] K. Choi, G. Fazekas, and M. Sandler, "Automatic tagging using deep convolutional neural networks," International Society for Music Information Retrieval Conference, pp. 805–811, 2016.

[6] A. Goodfellow, Ian, Bengio, Yoshua, Courville, "Deep Learning," MIT Press, 2016.

[7] J. Riionheimo and V. Välimäki, "Parameter estimation of a plucked string synthesis model using a genetic algorithm with perceptual fitness calculation," Eurasip Journal on Applied Signal Processing, vol. 2003, no. 8, pp. 791–805, 2003.

[8] A. W. Y. Su and S. F. Liang, "A class of physical modeling recurrent networks for analysis/synthesis of plucked string instruments," IEEE Transactions on Neural Networks, vol. 13, no. 5, pp. 1137–1148, 2002.

[9] M. Sterling and M. Bocko, "Empirical physical modeling for bowed string instruments," in ICASSP, IEEE International Conference on Acoustics, Speech and Signal Processing - Proceedings, 2010, pp. 433–436.

[10] K. Itoyama and H. G. Okuno, "Parameter Estimation of Virtual Musical Instrument Synthesizers," in Proceedings of ICMC, 2014, pp. 95–101.

[11] http://www.softsynth.com/jsyn

[12] Van Den Oord Aaron and K. Kavukcuoglu, "WaveNet: A Generative Model for Raw Audio," arXiv preprint arXiv:160903499, 2016.

[13] C. M. Bishop, Pattern Recognition and Machine Learning, vol. 4, no. 4. 2006.

[14] D. P. Kingma and J. L. Ba, "Adam: a Method for Stochastic Optimization," International Conference on Learning Representations 2015, pp. 1–15, 2015.

[15] https://github.com/deepsynth/deepsynth

[16] A. Krizhevsky, I. Sutskever, and G. E. Hinton, "ImageNet Classification with Deep Convolutional Neural Networks," in Advances in Neural Information Processing Systems 25, 2012, pp. 1097–1105.

[17] A. Graves, A. Mohamed, and G. E. Hinton, "Speech recognition with deep recurrent neural networks," in {IEEE} International Conference on Acoustics, Speech and Signal Processing, {ICASSP} 2013, Vancouver, BC, Canada, May 26-31, 2013, 2013, pp. 6645–6649.

[18] Y. Bengio, R. Ducharme, P. Vincent, and C. Janvin, "A Neural Probabilistic Language Model," J. Mach. Learn. Res., vol. 3, pp. 1137–1155, Mar. 2003.

[19] W. Yih, K. Toutanova, J. C. Platt, and C. Meek, "Learning Discriminative Projections for Text Similarity Measures," in Proceedings of the Fifteenth Conference on Computational Natural Language Learning, 2011, pp. 247–256.

[20] T. Mikolov, I. Sutskever, K. Chen, G. S. Corrado, and J. Dean, "Distributed Representations of Words and Phrases and their Compositionality," in Advances in Neural Information Processing Systems 26, 2013

[21] R. Collobert, J. Weston, L. Bottou, M. Karlen, K. Kavukcuoglu, and P. Kuksa, "Natural Language Processing (Almost) from Scratch," J. Mach. Learn. Res., vol. 12, pp. 2493–2537, Nov. 2011.

[22] Y. Kim, "Convolutional Neural Networks for Sentence Classification," Proc. 2014 Conf. Empir. Methods Nat. Lang. Process. (EMNLP 2014), pp. 1746–1751, 2014.

[23] Barkan, O. Bayesian Neural Word Embedding. AAAI 2017.

[24] He K, Zhang X, Ren S, Sun J. Deep residual learning for image recognition. In Proceedings of the IEEE conference on computer vision and pattern recognition 2016 (pp. 770-778).

[25] Huang G, Liu Z, Van Der Maaten L, Weinberger KQ. Densely Connected Convolutional Networks. In CVPR 2017 Jul 21 (Vol. 1, No. 2, p. 3).

[26] Barkan O, Tsiris D. Deep Synthesizer Parameter Estimation. ICASSP 2019.

[27] M. Yee-King et al., "Automatic Programming of VST Sound Synthesizers Using Deep Networks and Other Techniques." IEEE Transactions on Emerging Topics in Computational Intelligence, 2018.

[28] M. Yee-King, "Automated Sound Synthesizer Programming: Techniques and Applications" PhD Thesis, University of Sussex, 2012.

[29] Hu, Yuanming et al., "Exposure: A white-box photo post-processing framework." ACM Transactions on Graphics (TOG)37.2, 2018.

[30] Sheng, Di, and György Fazekas. "A Feature Learning Siamese Model for Intelligent Control of the Dynamic Range Compressor." arXiv preprint arXiv:1905.01022, 2019.

[31] Sheng, Di, and György Fazekas. "Automatic control of the dynamic range compressor using a regression model and a reference sound." Proceedings of the 20th International Conference on Digital Audio Effects (DAFx-17), 2017.

[32] Jacovi, Alon, et al. "Neural network gradient-based learning of black-box function interfaces." arXiv preprint arXiv:1901.03995, 2019.

[33] Yan, Zhicheng, et al. "Automatic photo adjustment using deep neural networks." ACM Transactions on Graphics (TOG) 35.2, 2016.

[34] E-P Damskagg, et al. "Deep Learning for Tube Amplifier Emulation." ICASSP, 2019.

[35] Martinez, Marco and Joshua D. Reiss. "Modeling of nonlinear audio effects with end-to-end deep neural networks." ICASSP, 2019.

[36] Martinez, Marco and Joshua D. Reiss. "End-to-end Equalization with Convolutional Neural Networks." DAFX, 2018.

[37] S. H. Hawley et al. "SignalTrain: Profiling Audio compressors with Deep Neural Networks." arXiv preprint arXiv:1905.11928, 2019.

[38] Katsigiannis S, Scovell J, Ramzan N, Janowski L, Corriveau P, Saad MA, Van Wallendael G. Interpreting MOS scores, when can users see a difference? Understanding user experience differences for photo quality. Quality and User Experience. 2018 Dec 1;3(1):6.

[39] https://ccrma.stanford.edu/~jos/pasp/

**Oren Barkan** biography is not available.

**David Tsiris** biography is not available.

**Noam Koenigstein** biography is not available.

**Ori Katz** biography is not available.